\documentclass[journal=jacsat,manuscript=article]{achemso}
\usepackage{amsmath}
\usepackage[T1]{fontenc} 
\usepackage[utf8]{inputenc}
\usepackage{textcomp}
\DeclareUnicodeCharacter{03B4}{$\delta$}
\DeclareUnicodeCharacter{03B1}{$\alpha$}

\def\SB#1{\textsubscript{{#1}}}

\author{Paramvir Ahlawat}
\email{paramvir.chem@gmail.com}

\title[] {Molecular dynamics simulations of nucleation of hexagonal($\delta$) and cubic($\alpha$)-FAPbI\textsubscript{3} perovskites from solution}

\abbreviations{}

\begin{document}

%
%
%
%
%

\begin{abstract}
Solar to power conversion certified efficiencies of formamidinium lead iodide\cite{karlin_synthesis_2007, stoumpos_semiconducting_2013, eperon_formamidinium_2014, lee_high-efficiency_2014, koh_formamidinium-containing_2014, pang_nh_2014, pellet_mixed-organic-cation_2014} based single-junction perovskite solar cell (PSC)\cite{kojima_organometal_2009, im_6.5_2011, lee_efficient_2012} is now 26\%\cite{zhao_inactive_2022, park_controlled_2023, NREL}, all perovskite-perovskite tandem 28\%\cite{chen_bifacial_2022, lin_all-perovskite_2023, he_improving_2023}, and the perovskite-silicon tandem solar cell is $\sim$34\%\cite{NREL, noauthor_leaders_nodate, sahli_fully_2018, jost_textured_2018, chen_blade-coated_2020, liu_efficient_2022} going beyond gallium arsenide solar cells\cite{NREL}. Therefore, it is now one of \textbf{the most promising materials for generating cheaper sunlight-based electricity}. This material has two commonly known polymorphs; one is a thermodynamically stable yellow hexagonal phase. The other one is a metastable black perovskite phase: a powerful photo-active material. Thousands of experiments have been performed to produce the highly crystalline photoactive phase of formamidinium lead iodide. Despite that, PSCs often suffer from poor reproducibility and stability. One of the root cause is the lack of control over their synthesis, \textit{i.e.} crystallization process, where \textbf{one of the main quests is to synthesize and stabilize corner-sharing defects-free photoactive perovskite form of pure or doped black perovskite form of formamidinium lead iodide and prevent the formation of hexagonal phases}. Thus, for the rapid industrialization of perovskite based solar farms to combat global rising temperatures: it is all-important to understand the polymorph selective nucleation of formamidinium lead iodide from its precursors. Towards this ultimate goal, here we perform molecular simulations of the nucleation of formamidinium lead iodide from solution. This study aims to take the primary steps for the all-atoms simulations of the polymorph selective crystallization of halide perovskites.
\end{abstract}

\section*{Main}
Solution processing\cite{green_emergence_2014, graetzel_materials_2012, park_perovskite_2015, berry_perovskite_2017, jena_halide_2019} for growing thin films of formamidinium lead iodide (FAPbI\textsubscript{3}) perovskite has been \textbf{one of the main driving forces for the relatively more straightforward synthesis of record efficiency perovskite solar cells (PSCs)}. In this process, countless experimental recipes have been tried to crystallize defects free black-FAPbI\textsubscript{3} by utilizing: (a) different process conditions\cite{burschka_sequential_2013, borchert_large-area_2017, liu_high-performance_2017, myung_challenges_2022, lu_vapor-assisted_2020, zhao_inactive_2022, sanchez_thermally_2022}, (b) numerous solvents\cite{jeon_solvent_2014, xiao_solvent_2014, yang_high-performance_2015, noel_low_2017, jeong_perovskite_2019, zhou_manipulating_2016, zhang_universal_2022, park_green_2020, kim_how_2021}, (c) additives\cite{abdi-jalebi_maximizing_2018, masi_chemi-structural_2020, wang_additive-modulated_2015, kim_methylammonium_2019, smith_layered_2014, lee_2d_2018, jeong_pseudo-halide_2021}, and (d) dopants\cite{lee_formamidinium_2015, saliba_incorporation_2016, duijnstee_understanding_2023, kim_impact_2020, caprioglio_open-circuit_2023, zhao_inactive_2022, lee_solid-phase_2020}. In the race for achieving higher efficiencies, endless experimental efforts are ongoing for synthesizing high crystal quality of phase-pure or alloyed corner-sharing perovskite photo-active structure of FAPbI\textsubscript{3}. However, a critical problem with the current experimental recipes is their reproducibility and scalability. One can only make >25\% PSCs on a tiny area\cite{NREL} of 0.1 - 1.0 cm\textsuperscript{2} and is not scalable enough to produce similar efficiency larger industrial scale PSC modules\cite{NREL_champion_nodate}. A complete atomic-level comprehensive picture of perovskites' crystallisation process can help to establish reproducible experimental methodologies for developing stable and record efficiency PSCs, especially large solar cell modules required for solar farms. However, it is of great challenge to extract the molecular details of a crystallization process. Now, \textit{state-of-the-art} experimental spectroscopic techniques are frequently limited by their spatio-temporal resolutions to seeing the complex dynamic process of nucleation. Over the years, transmission electron microscopy (TEM)\cite{nakamuro_capturing_2021, cao_atomic_2020, nielsen_situ_2014, loh_multistep_2017, li_direction-specific_2012} have been extensively employed to extract the atomic-level details of nucleation. However, halide perovskites are soft materials and sensitive to electron beams which can induce \textit{in-situ} degradation and subsequent phase changes during spectroscopic experiments; for example, metastable phases can convert to stable ones. Furthermore, it is a huge challenge to design electron microscopy experiments to study the formation of metastable phases of halide perovskite that are usually crystallized at elevated temperatures and from mixtures of complex solvents such as dimethylformamide (DMF), dimethyl sulfoxide (DMSO), gamma-butyrolactone (GBL), tetrahydrofuran (THF), 2-methoxy ethanol (2-ME) and a wide range of additives. Here, molecular simulations\cite{anwar_uncovering_2011, piana_simulating_2005, grunwald_nucleation_2009, chakraborty_how_2013, espinosa_seeding_2016, goswami_thermodynamics_2021, olenius_free_2013, vehkamaki_classical_2006, kawasaki_formation_2010, radu_enhanced_2017, auer_prediction_2001, salvalaglio_molecular-dynamics_2015, zimmermann_nucleation_2015, binder_overview:_2016, binder_theory_1987} can help in understanding the polymorph selective crystallization of FAPbI\textsubscript{3} and be of great use for making stable and reproducible solar cells. However, there are also challenges to performing MD simulations of the nucleation of this complex system from solutions. First, a typical simulation of nucleation from solution contains thousands of atoms/molecules, making it very difficult to simulate at the \textit{first-principles} level; for example, with the density functional theory (DFT). Secondly, nucleation processes are often rare event\cite{peters_reaction_2017} characterized by high free energy barriers, and it is computationally demanding to simulate experimental time scales with conventional computational architectures. In the third place, we could also face finite-size effects while simulating nucleation from solution. In this work, we take elementary steps to solve these ambitious problems and indeed, for the \textbf{first time successfully perform all-atoms nucleation of both hexagonal and cubic phases of FAPbI\textsubscript{3} from their precursor solutions}. First a scaled point charge inter-atomic potential is constructed for FAPbI\textsubscript{3} of a common functional form as AMBER\cite{AMBER_FF}, CHARMM\cite{CHARM_FF}, OPLS\cite{OPLS_FF}, and GROMOS\cite{GROMOS_FF}, see Equation \ref{eqn:LJ}. This potential is inspired by previous research work of Vega \textit{et al.} \cite{Vega_1, Vega_2, Vega_3} where scaled charge inter-atomic potentials are shown to reproduce various experimental quantities of ions in solutions.

\begin{equation}\label{eqn:LJ}
  \begin{aligned}
       u(r_{ij}) &= \sum_{electrostatics} \frac{1}{4\pi\epsilon_0} \frac{q_iq_j}{r_{ij}} + \sum_{Lennard-Jones} 4\epsilon_{ij}\left[ \left( \frac{\sigma_{ij}}{r_{ij}} \right)^{12} -  \left( \frac{\sigma_{ij}}{r_{ij}} \right)^{6} \right] + \\
                & \sum_{bonds} \frac{1}{2} k_b (\textbf{r}-\textbf{r}_0) + \sum_{angles} \frac{1}{2} k_{\theta} (\theta-{\theta}_0) + \sum_{torsions}\sum_n \frac{V_n}{2} \left( 1 + cos(n\Phi - \delta \right) \\
  \end{aligned}
\end{equation}



\begin{figure}[H]
  \includegraphics[width=150mm]{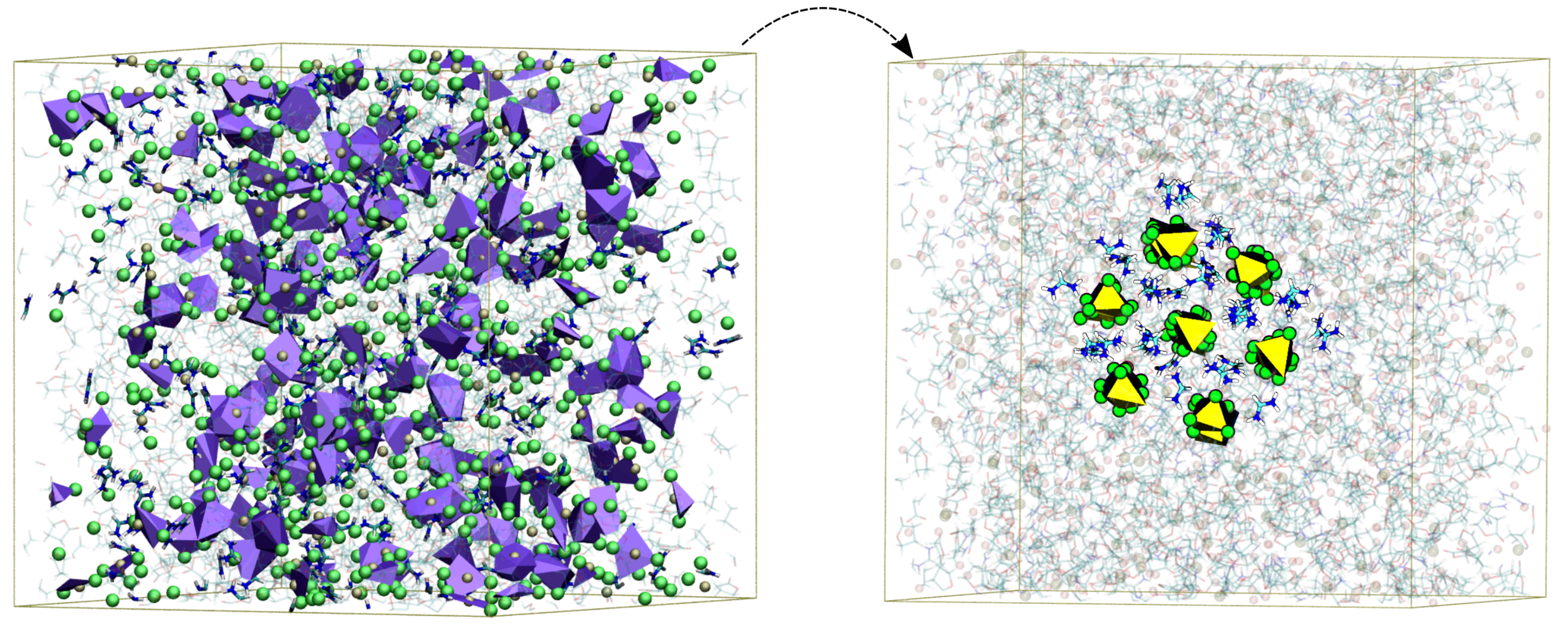}
  \label{fig:1}
  \caption{displays the nucleated hexagonal face-sharing yellow hexagonal phase of FAPbI\textsubscript{3} from its precursor solution}
\end{figure}

\begin{figure}[H]
  \includegraphics[width=150mm]{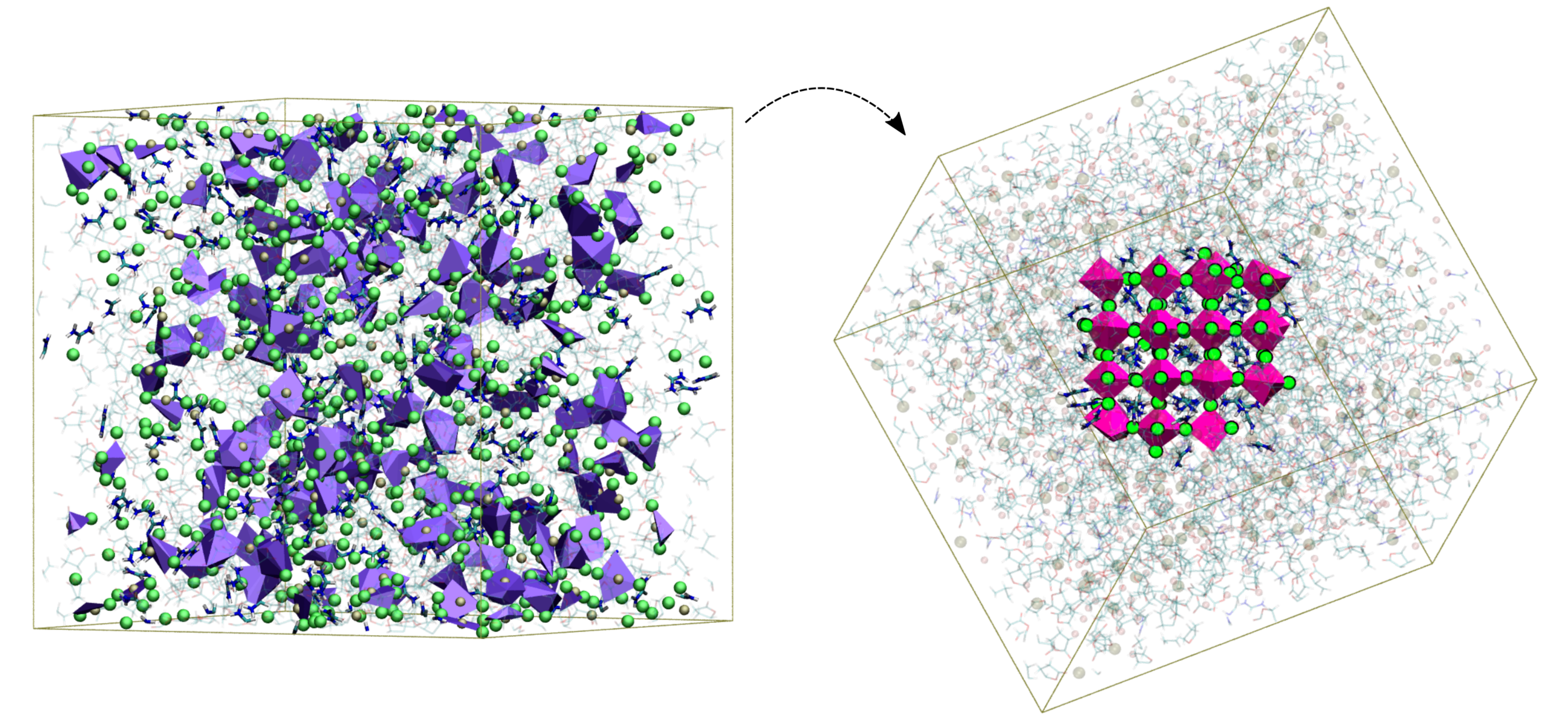}
  \label{fig:2}
  \caption{displays the nucleated corner-sharing perovskite phase of FAPbI\textsubscript{3} from its precursor solution}
\end{figure}

Now, to overcome the time scale problem, we take inspiration from the earlier studies of nucleation \textit{et al.}\cite{Daan_2, van_duijneveldt_computer_1992} and employ biased enhanced sampling techniques\cite{torrie_monte_1974, laio_escaping_2002, maragakis_gaussian-mixture_2009, invernizzi_opes:_2021}, where a bias potential explore the free energy surface as a function of system coordinates. One of the key challenge to perform the biased simulations is to make a suitable functional form of collective coordinates, commonly known as collective variables (CVs). \textbf{In this work, we introduce a CV that can be exploited to perform nucleation of any perovskite or in general to multi-species systems for example atmospheric aerosol particle nucleation\cite{vehkamaki_thermodynamics_2012, kulmala_direct_2013, kulmala_formation_2004} which is crucial to understand global climate change}. Our CV is based on combinations of the local coordination/density\cite{ornstein1914accidental, percus_analysis_1958, wolde_enhancement_1997, bartok_representing_2013, ahlawat_atomistic_2020} with the local crystalline order\cite{frenkel_structure_1986, ashcroft_structure_1967, lechner_accurate_2008, ahlawat_combined_2021} detailed in the following sections. First, we calculate the local densities ($\rho^{\alpha}(r_i^{Pb})$) of individual species (${\alpha}$) FA, Pb and I around the center position \textit{$r_i^{Pb}$} of Pb atoms:
\begin{equation}\label{eqn:CV11}
\rho^{\alpha}(r_i^{Pb}) = \frac{ 1 - \left(\frac{f_i^{\alpha}}{f^{\alpha}_{cut}}\right)^{-m}} 
{1 - \left(\frac{f_i^{\alpha}}{f^{\alpha}_{cut}}\right)^{-n}}
\end{equation}

Where $f_i^{\alpha}$ are the coordination numbers of FA, Pb and I around origin \textit{$r_i^{Pb}$} of Pb atoms defined as:

\begin{equation}\label{eqn:CV12}
f_i^{\alpha} = \sum_{j=1}^{N_{\alpha}} \frac{ 1 - \left(\frac{r_{ij}}{r^{\alpha}_{cut}}\right)^a } 
{ 1 - \left(\frac{r_{ij}}{r^{\alpha}_{cut}}\right)^b}
\end{equation}

Here $r_{ij}$ is the distance between positions of center Pb atoms \textbf{\textit{$r_i^{Pb}$}} and neighbouring species \textbf{\textit{$r_i^{\alpha}$}}(${\alpha}$ = FA, Pb and I). The values for $r^{\alpha}_{cut}$ and $f^{\alpha}_{cut}$ are chosen similar to the first coordination sphere of Pb atoms in their corresponding crystalline polymoprh of FAPbI\textsubscript{3}. Now to incorporate the local order, we calculate the local structure factors \cite{Debye_1, Debye_2, bonati_silicon_2018, niu_temperature_2019} ($s^{\alpha}(r_i^{Pb})$) of the species (${\alpha}$) FA, Pb and I around the center position \textit{$r_i^{Pb}$} of the Pb atoms. 

\begin{equation}\label{eqn:CV13}
s^{\alpha}(r_i^{Pb}) = \frac{ 1 - \left(\frac{s(q)_i^{\alpha}}{s(q)^{\alpha}_{cut}}\right)^{-c}} 
{1 - \left(\frac{s(q)_i^{\alpha}}{s(q)^{\alpha}_{cut}}\right)^{-d}}
\end{equation}

Where $s(q)_i^{\alpha}$ is defined as:

\begin{equation}\label{eqn:CV14}
s(q)_i^{\alpha} = 1 + \frac{1}{N}\sum_{j=i}^{N_{\alpha}} \frac{sin(qr\SB{ij})}{qr\SB{ij}}
\end{equation}

$N_{\alpha}$ are the number of species within similar distance cutoff used for calculating the coordination number $r^{\alpha}_{cut}$. The final local symmetric order centering Pb positions \textit{$r_i^{Pb}$} is calculated by multiplying\cite{percus_analysis_1958, ornstein1914accidental, van_leeuwen_new_1959} these individual local contributions from different species:

\begin{equation}\label{eqn:CV15}
\xi_i=\prod_{\alpha = FA, Pb, I} \rho^{\alpha}(r_i^{Pb}) s^{\alpha}(r_i^{Pb})
\end{equation}

The final expression for the CV becomes the sum for all Pb species:

\begin{equation}\label{eqn:CV16}
S_{p}= \sum_{i=1}^{N_{Pb}} \xi_i
\end{equation}

Where $N_{Pb}$ is the number of Pb ions around which the local order is calculated. We apply this functional form to selectively bias and nucleate the face-sharing and corner-sharing polymorphs of FAPbI\textsubscript{3} from their precursor solution. This CV is implemented in PLUMED-2.8\cite{tribello_plumed_2014, bonomi_plumed:_2009, the_plumed_consortium_promoting_2019}. We also test this implementation on other perovskite systems and, indeed observe nucleation from their homogeneous solution mixtures. Moreover, the idea of enhancing local atomic density along with the local crystalline order can be further extended to design more computationally efficient CVs.

\newpage

\begin{acknowledgement}
This research is funded by Swiss National Science Foundation (SNSF) through Postdoc.Mobility fellowship $P500PN\_206693$. 
\end{acknowledgement}

%
%


\section{Note}
All necessary data (input files, jupyter notebooks, other codes) to reproduce the results of this study are either deposited on Zenodo or can be accessed via github. 

\bibliography{achemso-demo}

\end{document}